\documentclass[prd,twocolumn,showpacs,nofootinbib,amsmath,amssymb,floatfix,superscriptaddress,showkeys]{revtex4}
\usepackage{multirow}
\usepackage{graphicx}
\usepackage{epstopdf}
\usepackage{dcolumn}
\usepackage{bm}
\usepackage{threeparttable}
\usepackage{subfigure}
\usepackage{color,txfonts}
\usepackage{ulem}
\definecolor{blue0}{rgb}{0,0,0.6}
\usepackage[colorlinks,linkcolor=blue0,anchorcolor=blue0,citecolor=blue0,urlcolor=blue0]{hyperref}

\newcommand{\jcap}{J. Cosmol. Astropart. Phys.}

\newcommand{\beq}{\begin{equation}}
\newcommand{\eeq}{\end{equation}}
\newcommand{\beqa}{\begin{eqnarray}}
\newcommand{\eeqa}{\end{eqnarray}}

\begin{document}

\title{Search for gamma-ray emission from the nearby dwarf spheroidal galaxies with 9 years of Fermi-LAT data}

\author{Shang Li}
\affiliation{Key Laboratory of Dark Matter and Space Astronomy, Purple Mountain Observatory, Chinese Academy of Sciences, Nanjing 210008, China}
\affiliation{University of Chinese Academy of Sciences, Yuquan Road 19, Beijing, 100049, China}
\author{Kai-Kai Duan}
\affiliation{Key Laboratory of Dark Matter and Space Astronomy, Purple Mountain Observatory, Chinese Academy of Sciences, Nanjing 210008, China}
\affiliation{University of Chinese Academy of Sciences, Yuquan Road 19, Beijing, 100049, China}
\author{Yun-Feng Liang}
\email{liangyf@pmo.ac.cn}
\affiliation{Key Laboratory of Dark Matter and Space Astronomy, Purple Mountain Observatory, Chinese Academy of Sciences, Nanjing 210008, China}

\author{Zi-Qing Xia}
\affiliation{Key Laboratory of Dark Matter and Space Astronomy, Purple Mountain Observatory, Chinese Academy of Sciences, Nanjing 210008, China}
\affiliation{School of Astronomy and Space Science, University of Science and Technology of China, Hefei, 230026, China}

\author{Zhao-Qiang Shen}
\email{zqshen@pmo.ac.cn}
\affiliation{Key Laboratory of Dark Matter and Space Astronomy, Purple Mountain Observatory, Chinese Academy of Sciences, Nanjing 210008, China}
\affiliation{University of Chinese Academy of Sciences, Yuquan Road 19, Beijing, 100049, China}
\author{Xiang Li}
\affiliation{Key Laboratory of Dark Matter and Space Astronomy, Purple Mountain Observatory, Chinese Academy of Sciences, Nanjing 210008, China}
\author{Neng-Hui Liao}
\affiliation{Key Laboratory of Dark Matter and Space Astronomy, Purple Mountain Observatory, Chinese Academy of Sciences, Nanjing 210008, China}
\author{Lei Feng}
\affiliation{Key Laboratory of Dark Matter and Space Astronomy, Purple Mountain Observatory, Chinese Academy of Sciences, Nanjing 210008, China}
\author{Qiang Yuan}
\affiliation{Key Laboratory of Dark Matter and Space Astronomy, Purple Mountain Observatory, Chinese Academy of Sciences, Nanjing 210008, China}
\affiliation{School of Astronomy and Space Science, University of Science and Technology of China, Hefei, 230026, China}
\author{Yi-Zhong Fan}
\email{yzfan@pmo.ac.cn}
\affiliation{Key Laboratory of Dark Matter and Space Astronomy, Purple Mountain Observatory, Chinese Academy of Sciences, Nanjing 210008, China}
\affiliation{School of Astronomy and Space Science, University of Science and Technology of China, Hefei, 230026, China}
\author{Jin Chang}
\affiliation{Key Laboratory of Dark Matter and Space Astronomy, Purple Mountain Observatory, Chinese Academy of Sciences, Nanjing 210008, China}

\date{\today}

\begin{abstract}
In this work, we search for $\gamma$-ray emission from the directions of some nearby Milky Way dwarf spheroidal galaxies (dSphs) and candidates with the publicly-available Pass 8 data of Fermi-LAT. Our sample includes 12 sources with the distances $<50$ kpc. Very weak $\gamma$-ray excesses ($\sim 2\sigma$) are found in some dSphs/candidates, consistent with that reported in the previous literature. Intriguingly, the peak test statistic (TS) value of the weak emission from the direction of Reticulum II rises continually. If interpreted as dark matter (DM) annihilation, the peak TS value is 13.5 for the annihilation channel of $\chi\chi \rightarrow \tau^{+}\tau^{-}$ and the DM mass of $m_\chi \sim 16$ GeV. The combination of all these nearby sources yields a more significant (with local significance $> 4\sigma$) $\gamma$-ray signal.

\end{abstract}

\pacs{95.35.+d, 95.85.Pw, 98.52.Wz}

\maketitle
\section{Introduction}
The latest observations suggest that non-baryonic cold dark matter (DM) constitutes $\sim$ 84\% of the matter density of the Universe \cite{Ade:2015xua}. The nature of DM particles is still a mystery at the moment. Among all the dark matter particle candidates, the weakly interacting massive particles (WIMPs) have attracted the most attention \cite{Jungman:1995df, Bertone:2004pz, Hooper:2007qk, Feng:2010gw}. WIMPs may annihilate or decay and finally produce GeV-TeV gamma-rays or cosmic rays.
The prime goal of the dark matter indirect detection is to recognize these products of dark matter origin, with experiments such as the Fermi Large Area Telescope (Fermi-LAT \cite{atwood09lat}) and the Dark Matter Particle Explorer (DAMPE \cite{DAMPE:2017,Chang:2017n}). Since the launch of Fermi satellite in 2008 \cite{atwood09lat}, many efforts have been made to search for gamma-ray signal from the DM annihilation or decay, but no reliable signal has been conclusively detected so far (see \cite{charles16review} for a recent review). 

Milky Way dwarf spheroidal (dSph) galaxy is considered as one of the most ideal targets for indirect detection of dark matter. Because the mass-to-light ratio of dSphs is predicted to be of the order of $10-10^{3}$ \cite{Wolf:2010,Simon:2011}, implying they are DM dominated. In addition, due to the lack of astrophysical activities \cite{Lake:1990du,Baltz:2004bb,Strigari:2013iaa}, the DM signal search with dSphs benefits from very low gamma-ray backgrounds. For a long time, no potential evidence of DM signal was found within the Fermi-LAT observations of dSphs, therefore very strong constraints on the DM parameters were obtained \cite{fermi11dsph,GeringerSameth:2011iw,2012PhRvD86b3528C,tsai13dsph,fermi14dsph,zhao2016ds,gs15dsph,fermi15dsph}.
In the past three years, thanks to several newly launched optical imaging surveys, more than 20 new dSphs and candidates have been found
\cite{des15y1, des15y2, laevens15tri2, laevens15_3, kim15pegasus3, Daisuke(2016), Daisuke(2017), Drlica-Wagner1(2016), Torrealba(2016)}.
Analyses on searching for the $\gamma$-ray emission from these newly discovered dSphs and candidates have also been conducted \cite{gs15ret2, hooper15ret2,Drlica-Wagner:2015xua,li16dsph,fermi2016dsph,liang2016dl}. Though none of these searches displays a statistically-significant signal,
evidence of very weak gamma-ray emission was reported in the Reticulum II \cite{gs15ret2, hooper15ret2,Drlica-Wagner:2015xua, fermi2016dsph} and Tucana III \cite{li16dsph, fermi2016dsph}. Intriguingly, the tentative emission is found to be consistent with the Galactic GeV excess reported in \cite{Hooper:2010mq, Abazajian:2012pn, Gordon:2013vta, Hooper:2013rwa, Daylan:2014rsa, Zhou:2014lva, Calore:2014xka, Huang:2015rlu, fermi17GCE}.

All these previous works are based on the Fermi-LAT data of less than 6 years.
In this paper, we analyze the 9 years of Fermi-LAT Pass 8 data to search for $\gamma$-ray emission from dSphs.
We focus on analyzing the nearby dSphs which are most promising to generate DM annihilation signal and testing whether the significance of the previously reported tentative emission from dSphs increases with the time or not.
Our sample is selected from the Table I of Ref. \cite{fermi2016dsph}, but limited to those with distances $<50$ kpc.

\begin{center}
\begin{table*}[!t]
\caption{Confirmed and Candidate dSphs with distances $<50\,{\rm kpc}$.}
\begin{tabular}{lcccccccc}
\hline
\hline
 Name &  $(l,b)$ & ${\rm Distance}$ & $\log_{10}{{\rm (J)}^{a}}$& $\log_{10}{{\rm (Est.\;J)}^{b}}$ & & ${\rm TS}_{\rm peak}^{b\bar{b}}$ & & ${\rm TS}_{\rm peak}^{\tau^{+}\tau^{-}}$ \\
     & [deg] &[kpc]&[$\log_{10}{\rm (GeV^{2}cm^{-5})}$] & [$\log_{10}{\rm (GeV^{2}cm^{-5})}$] & &  & & \\
\hline
Bootes II      &(353.69, 68.87) & $42$ &$-$& 18.9  & &    2.8  & &   2.9  \\
Bootes III     &(35.41, 75.35) & $47$ &$-$& 18.8   & &    4.3  & &   0    \\
Coma Berenices &(241.89, 83.61) & $44$ &19.0$\pm0.4$& 18.8  & &    0.4  & &   0.6  \\
Draco II       &(98.29, 42.88) & $24$ &$-$& 19.3   & &    1.8  & &   2.4  \\
Cetus II       &(156.47, -78.53) & $30$ &$-$& 19.1 & &    3.8  & &   2.8  \\
Reticulum II   &(266.30, -49.74) & $32$ &18.9$\pm0.6$& 19.1 & &   10.9  & &   13.5 \\
Segue 1        &(220.48, 50.43) & $23$ &19.4$\pm0.3$& 19.4  & &    0.6  & &   0.9  \\
Triangulum II  &(140.90, -23.82) & $30$ &$-$& 19.1 & &    0.8  & &   0.6  \\
Tucana III     &(315.38, -56.18) & $25$ &$-$& 19.3 & &    3.6  & &   4.6  \\
Tucana IV      &(313.29, -55.29) & $48$ &$-$& 18.7 & &    2.9  & &   2.9  \\
Ursa Major II  &(152.46, 37.44) & $32$ &19.4$\pm0.4$& 19.1  & &    1.7  & &   0.4  \\
Willman 1      &(158.58, 56.78) & $38$ & $-$&18.9  & &    2.6  & &   3.0  \\
\hline
\end{tabular}
\begin{tablenotes}
\item $^{\rm a}$ J-factors derived through stellar kinematics. The Reticulum II one is taken from \cite{Simon:2015fdw}, others are from \cite{Geringer-Sameth:2014yza}.
\item $^{\rm b}$ J-factors estimated with the empirical relation $J(d)\approx 10^{18.1\pm 0.1}(d/100~{\rm kpc})^{-2}$, where $d$ is the distance \cite{Drlica-Wagner:2015xua}.
\end{tablenotes}
\label{tab:12dsph}
\end{table*}
\end{center}

\section{Fermi-LAT data analysis}
In this paper, we use the nine-year (2008 August 4 to 2017 August 4) Fermi-LAT Pass 8 data in the energy range from 500 MeV to 500 GeV. In order to remove the $\gamma$-rays produced by cosmic-ray interactions in the Earth's atmosphere, the photons with zenith angle greater than $100^{\circ}$ are rejected. We use the quality-filter cuts ({\tt DATA\_QUAL==1 \&\& LAT\_CONFIG==1}) to ensure the data are valid for science use. The latest analysis software {\tt Fermi Science Tools} of version {\tt v10r0p5} is applied to analyze the Fermi-LAT data. We take a $5^{\circ}$ region of interest (ROI) for each dSph/candidate (hereafter we use the word {\it dSph} to denote both confirmed and candidate dSphs), which is sufficient to include all the $> 500\,{\rm MeV}$ photons from the target considering the angular resolution of the Fermi-LAT data.
The initial background models are generated by the script {\tt makeFL8Yxml.py}\footnote{\url{https://fermi.gsfc.nasa.gov/ssc/data/access/lat/fl8y/makeFL8Yxml.py}}, with all FL8Y\footnote{\url{https://fermi.gsfc.nasa.gov/ssc/data/access/lat/fl8y/gll_psc_8year_v3.fit}} sources within 10$^{\circ}$ around each target as well as the latest diffuse gamma-ray emission models {\tt gll\_iem\_v06.fits} and {\tt iso\_P8R2\_SOURCE\_V6\_v06.txt} included. 
Similar to Ref. \cite{fermi15dsph}, we model the dSph candidates with spatially extended Navarro-Frenk-White (NFW) DM density profiles \cite{Navarro:1996gj}. For Coma Berenices, Segue 1 and Ursa Major II, we use the scale radii reported in \cite{Geringer-Sameth:2014yza}; while the scale radii of other dSphs are set to 1 kpc.
For some analyses, we also present the results of using point source to model the target. The variation between the two sets of results gives indications of how the spatial model affecting the final results.

\begin{figure*}[t]
\begin{center}
\includegraphics[width=1.0\textwidth]{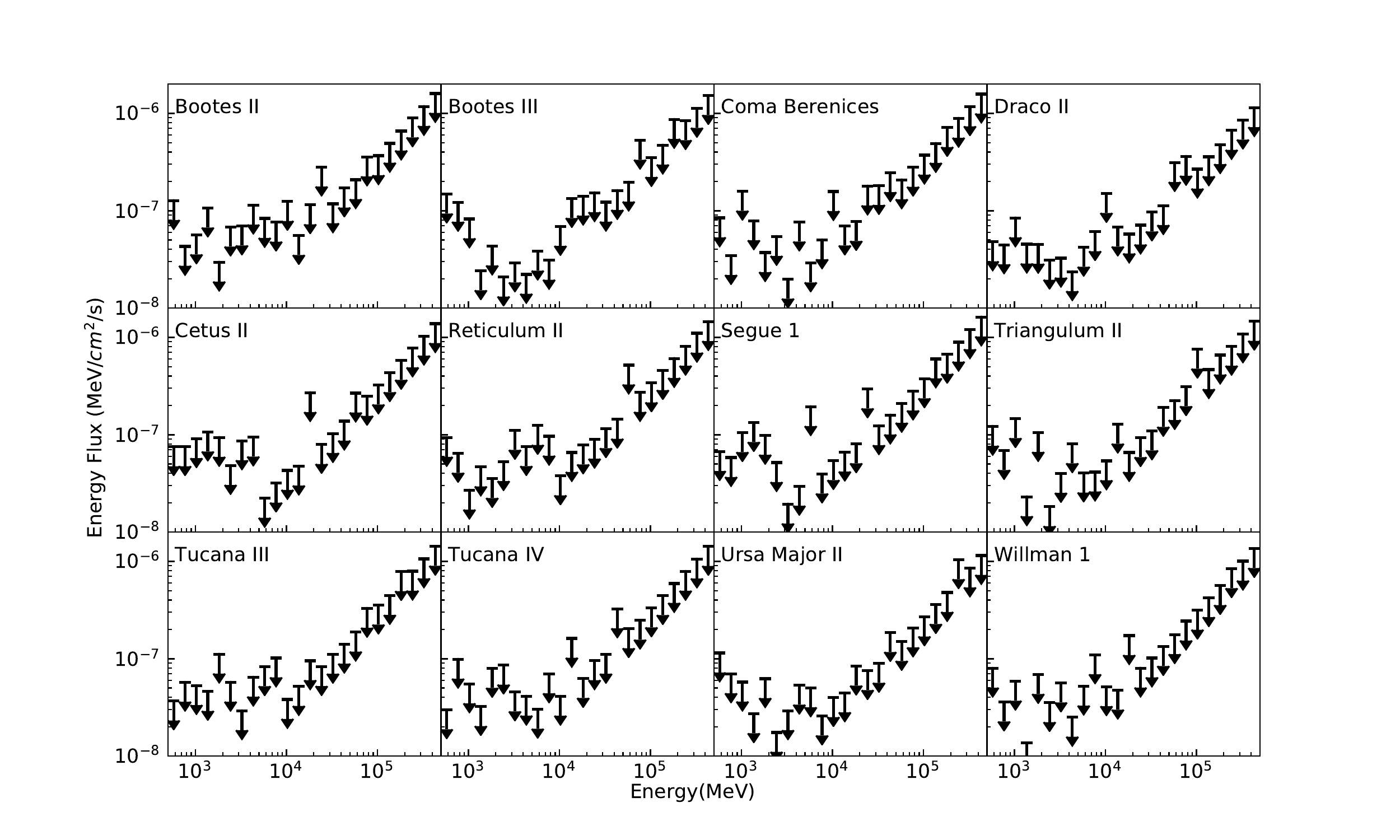}
\end{center}
\caption{Bin-by-bin integrated energy-flux upper limits at 95\% confidence level assuming a point-like model for the 12 targets.}
\label{Fig.1}
\end{figure*}

We first adopt a standard unbinned likelihood analysis\footnote{\url{https://fermi.gsfc.nasa.gov/ssc/data/analysis/scitools/likelihood_tutorial.html}} to derive the best-fit parameters for the background sources (nuisance parameters). In this procedure, the prefactors of all the point sources within 5$^{\circ}$ ROI together with the normalizations of two diffuse backgrounds are set free in the likelihood fitting.
Later, a bin-by-bin likelihood profile method is utilized to obtain the TS value and flux upper limit of the tentative $\gamma$-ray emission associated with each target. The likelihood profile method will facilitate the scanning of a series of DM masses and the combined likelihood analysis in Sec. \ref{sec:combined}.
The entire data set in the energy range from 500 MeV to 500 GeV are divided into 24 logarithmically spaced energy bins.
For each energy bin $k$, we model the putative dSph source by a power-law spectrum ($dN/dE \propto E^{-\Gamma}$) with $\Gamma$=2 \cite{fermi14dsph,fermi15dsph} to derive the relation between the likelihood $L_k$ and the target's flux $f_k$ (i.e., likelihood profile). The flux in a narrow energy bin is found independent of the choice of spectral model. To get better sensitivity, we also perform an unbinned likelihood analysis when generating the likelihood profile. In Fig.~\ref{Fig.1} we show the bin-by-bin integrated energy-flux upper limits at 95\% confidence level for the 12 dSphs.
A broadband likelihood function is constructed by multiplying the bin-by-bin likelihoods evaluated at the predicted fluxes for a given spectral model,
\begin{equation}
L(\bm{\alpha},\bm{\theta})=\prod_k L_k(f_k(\bm{\alpha}),\bm{\theta}),
\label{eq1}
\end{equation}
where $\bm{\alpha}$ and $\bm{\theta}$ are target and nuisance parameters, respectively. 
We take the similar method as developed in \cite{fermi11dsph,tsai13dsph,fermi14dsph}, except that we adopt an unbinned analysis and consider longer data set. We refer readers to these literatures for more details of the method.

\begin{figure*}[t]
\includegraphics[width=0.45\textwidth]{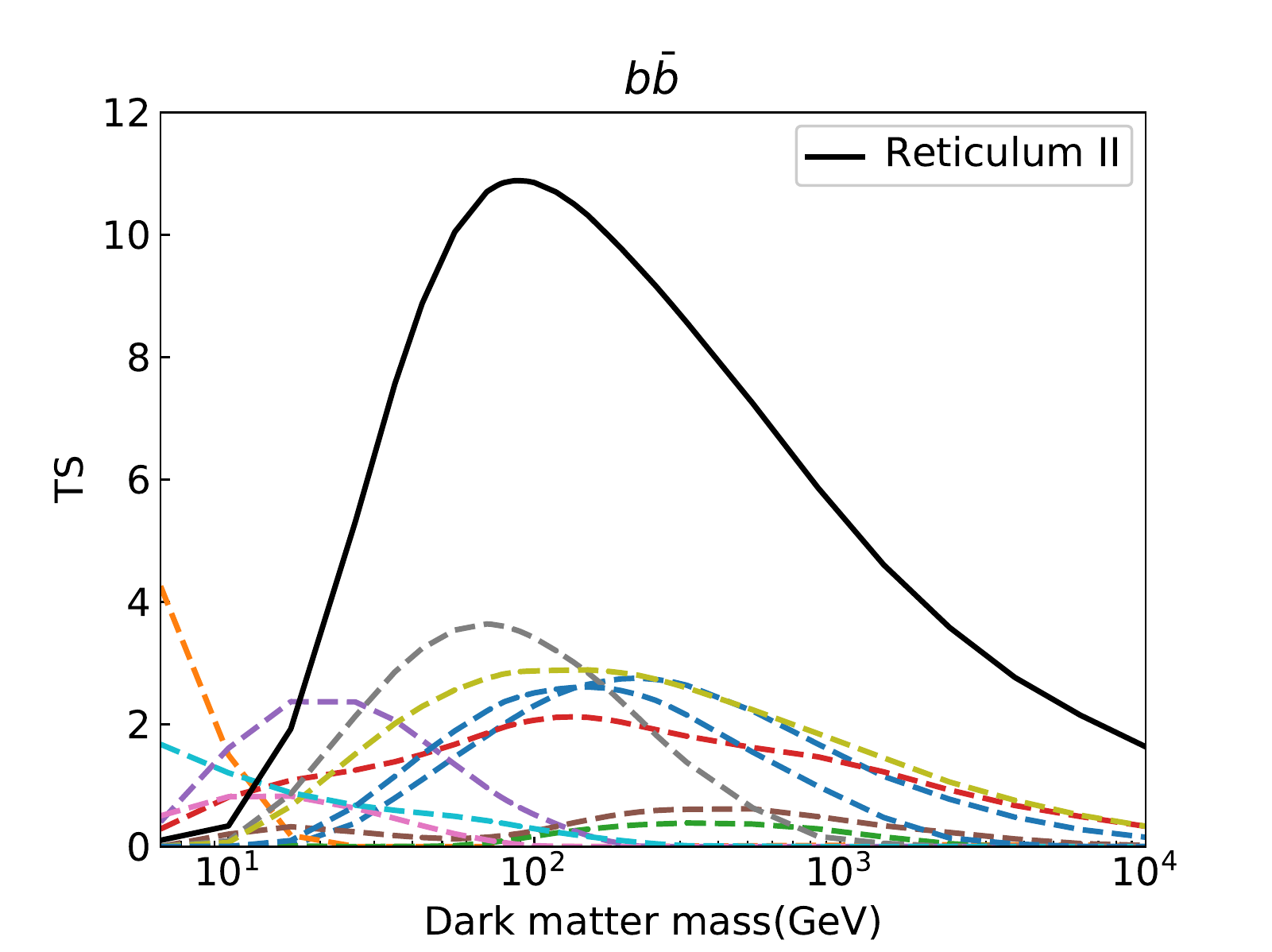}
\includegraphics[width=0.45\textwidth]{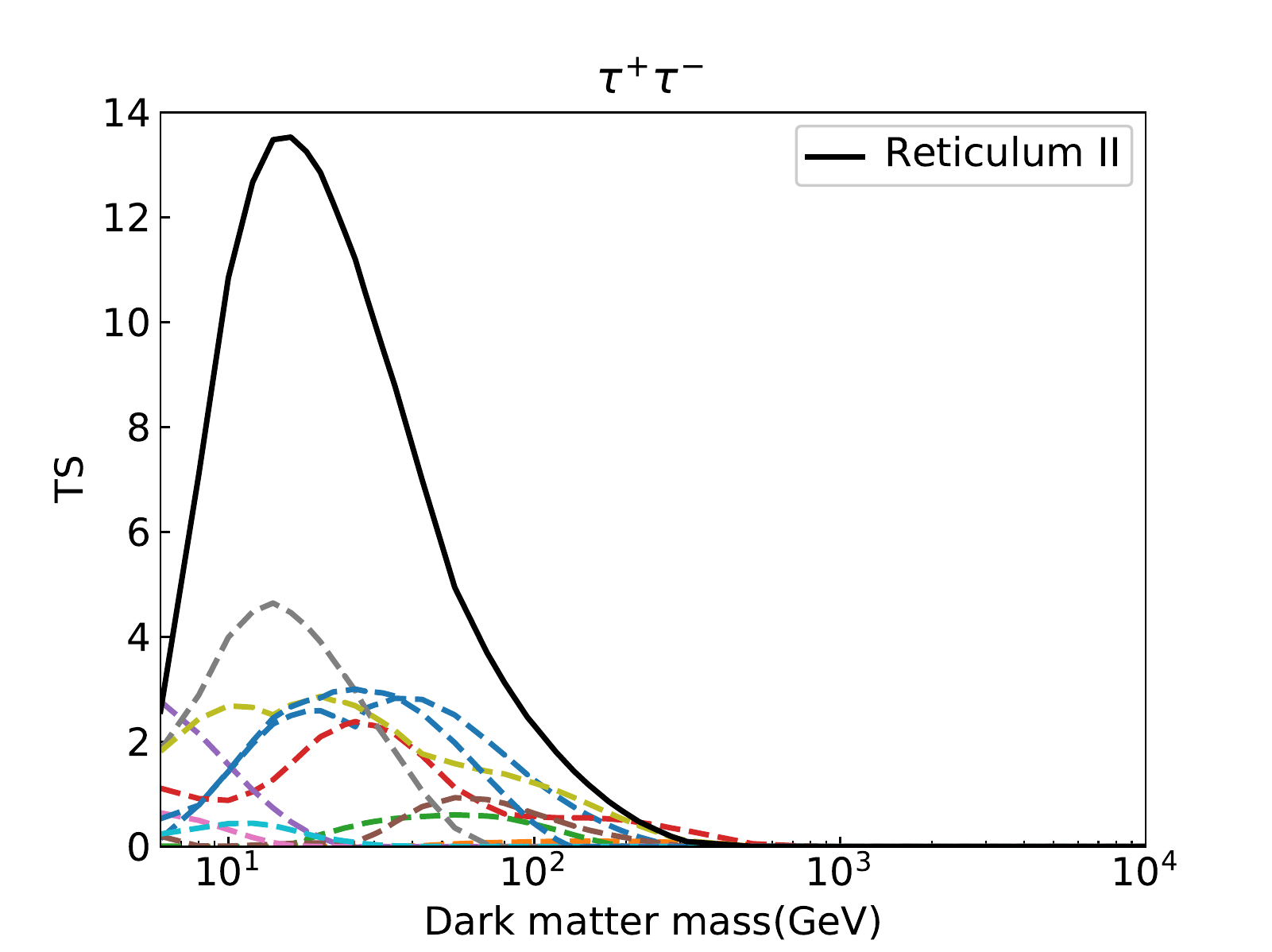}
\caption{Local detection significance, expressed as a log-likelihood test statistic (TS), as a function of the DM particle mass for the 12 dSphs listed in Table \ref{tab:12dsph} assuming DM annihilating through the $b\bar{b}$ (left) or ${\tau^{+}\tau^{-}}$ (right) channels.}

\label{fig:ts}
\end{figure*}

\section{Searching for dark matter emission from dSphs}
DSph is one of the most ideal targets for indirect detection of dark matter. The ${\gamma}$-ray flux expected from annihilation of DM particles is expressed as \cite{Jungman:1995df, Bertone:2004pz, Hooper:2007qk, Feng:2010gw}
\begin{equation}
{\Phi}(E_{\gamma})={\frac{\left<{\sigma}v\right>}{8{\pi}m_{\chi}^{2}}\frac{dN_{\gamma}}{dE_{\gamma}}\times J},
\end{equation}
where ${m_{\chi}}$, ${\left<{\sigma}v\right>}$, $dN_{\gamma}/dE_{\gamma}$ are the rest mass, thermal averaged annihilation cross section and the differential gamma-ray production per annihilation of the dark matter particle. The term $$J={\int}{\rho}^{2}(r)dld{\Omega}$$ is the line-of-sight integral of the square of the dark matter density (i.e., the so-called J-factor). In this work, the DM spectra are obtained with PPP4DMID \cite{Cirelli:2010xx}.

\subsection{Individual study}
In this section, we search for the excess $\gamma$-ray emission from each of the 12 selected dSphs. We consider two representative DM annihilation channels (i.e., ${b\bar{b}}$ and ${\tau^{+}\tau^{-}}$), and scan the DM particle masses from 6 GeV to 10 TeV. The significance of the target source can be quantified with the test statistic (TS), which is expressed as ${\rm TS}=-2{\rm ln}(L_{0}/L)$, where $L$ and $L_{0}$ are the maximum likelihood values for the models with and without the putative dSph source, respectively. Please note that the TS value of an individual target is independent of the given J-factor and its uncertainty.

In Fig.~\ref{fig:ts} we show the TS values of $\gamma$-ray signal from our sample for two representative annihilation channels and a series of dark matter masses. From Fig.~\ref{fig:ts}, we can see that no dSph shows significant (i.e., $\rm TS>25$) $\gamma$-ray signal even though much more data are considered in our work. 
The most significant signal appears in the direction of Reticulum II. For this dSph, in the case of $\chi\chi\rightarrow\tau^{+}\tau^{-}$, the TS value of the fit peaks at $m_\chi \sim 16$ GeV with ${\rm TS_{peak}}\sim13.5$ (12.2). For $\chi\chi\rightarrow b\bar{b}$, the largest TS value is $\sim10.9$ (10.0) corresponding to the DM mass of $m_\chi \sim 90$ GeV. The TS values in the brackets are obtained by using a point source to model Reticulum II.

Some previous works have also analyzed the gamma-ray emission from Reticulum II \cite{gs15ret2,hooper15ret2,Drlica-Wagner:2015xua,fermi2016dsph}.
Shortly after the DES Collaboration released their first group of dSph candidates, evidence of $\gamma$-ray emission from Reticulum II was reported based on the analysis of Fermi-LAT P7REP data \cite{gs15ret2,hooper15ret2}. 
However, the subsequent studies on this source with updated Pass 8 data of Fermi-LAT resulted in lower significances \cite{Drlica-Wagner:2015xua,fermi2016dsph}, and hence these authors did not claim excess from Reticulum II.
A summary of previous analysis results are presented in Table \ref{tab:ret2}. Focusing on the Pass 8 analyses, we find that our results here give the highest TS values. Since the analysis approach adopted in this work is considerably similar to the previous ones, the increase of the TS value is most possibly owing to the longer data used. Such a behavior of the excess suggests it resembles a real signal, no matter an astrophysical or dark matter origin.

\begin{figure}[!h]
\includegraphics[width=0.9\columnwidth]{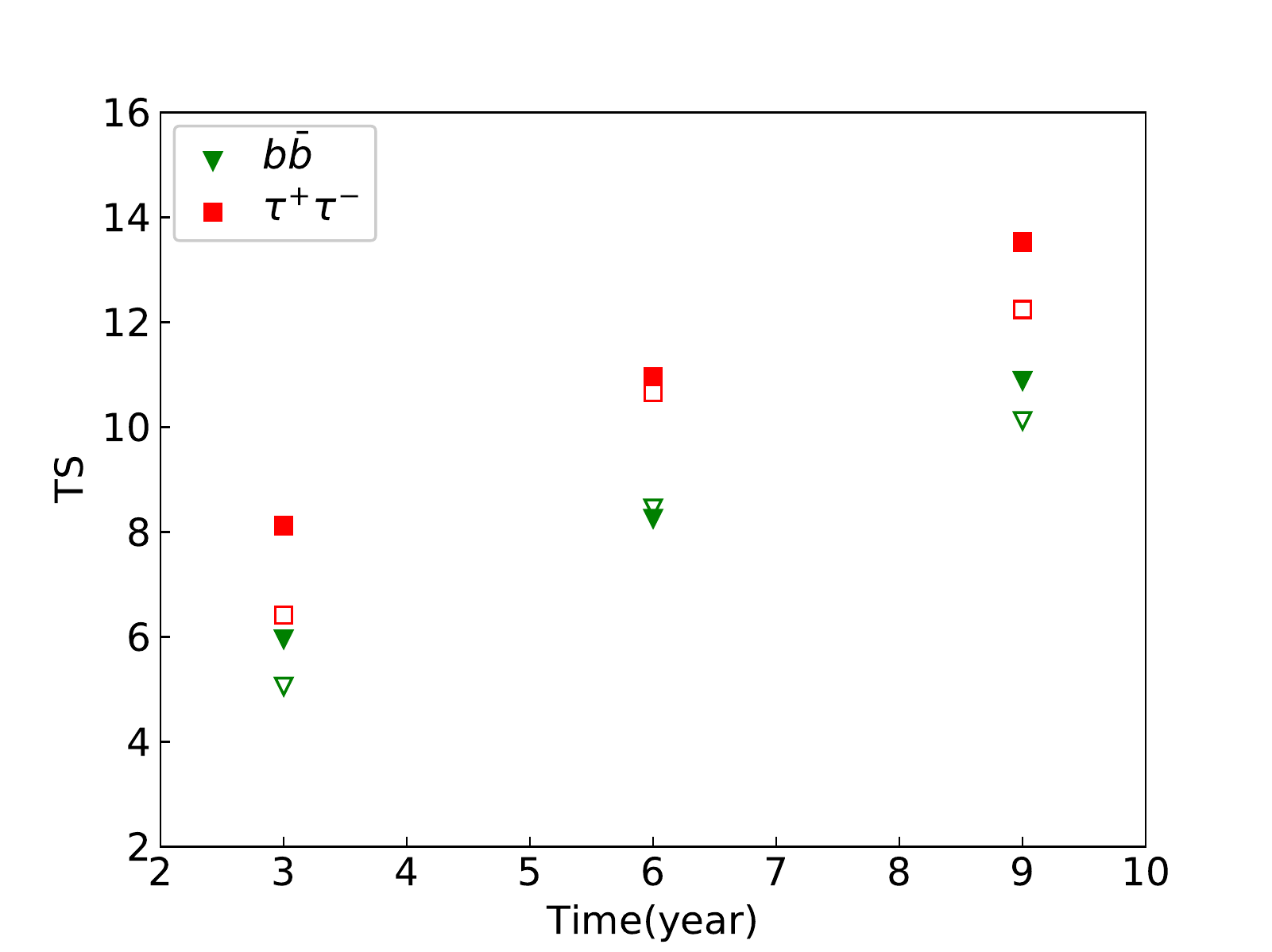}
\caption{The $\rm TS_{peak}$ of the $\gamma$-ray emission in the direction of Reticulum II for $b\bar{b}$ and ${\tau^{+}\tau^{-}}$ annihilation channels in three different time intervals. The open points represent the results of using a point source to model the dSph.}
\label{fig:3interval}
\end{figure}

The $\gamma$-ray excess from Reticulum II seems encouraging. Here we further test whether the TS value of the excess increases with the data accumulation.
We apply the same analysis procedure to derive the TS values of the potential excess for the Fermi-LAT data of 3, 6 and 9 years. In Fig.~\ref{fig:3interval}, we present the peak TS values of Retucilum II for three time intervals considered. As is shown, the peak TS value is indeed increasing with time for both annihilation channels. Such a trend is expected in the models of dark matter annihilation or alternatively a steady astrophysical source. Although the results here seem interesting, we should realize that the {\it signal} now is too weak to rule out the possibility of a statistical fluctuation origin. More observations to this target are needed to draw a robust conclusion.


\begin{center}
\begin{table}[!h]
\caption{Summary of analysis results of Reticulum II.}
\begin{tabular}{lccccccc}
\hline
\hline
 Results$^{a}$          & Data Size & Data Set & $\sigma_{b\bar{b}}$ &Mass & $\sigma_{\tau^{+}\tau^{-}}$&Mass \\
            & [years] &  &  & [GeV] & & [GeV] \\
\hline
\cite{gs15ret2}     & 6.5 & P7REP & $4.1$ & $\sim75$   &     $4.3$  &  $\sim14$    \\
\cite{hooper15ret2} & 6.5 & P7REP & $4.2$ & 49  &     $-$  &   $-$  \\
\cite{Drlica-Wagner:2015xua}  & 6 & Pass 8 & $-$ & $-$  &     2.6  &   25  \\
\cite{fermi2016dsph} &  6 & Pass 8 & 2.4 & 100   &     2.6  &    15.8  \\
This work              & 9  & Pass 8 & 3.3 & 90 &    3.7  &    16  \\
\hline
\end{tabular}
\begin{tablenotes}
\item $^{\rm a}$ The results are reported by Geringer-Sameth et al. \cite{gs15ret2}, Hooper \& Linden \cite{hooper15ret2}, Fermi+DES Col. \cite{Drlica-Wagner:2015xua} and Fermi Col. \cite{fermi2016dsph}. 
\item $^{\rm b}$ The 4th and 6th columns are the local detection significances for the tentative excess in the direction of Reticulum II.
\end{tablenotes}
\label{tab:ret2}
\end{table}
\end{center}

\subsection{Combined analysis}
\label{sec:combined}

Under the assumption that the properties of dark matter particles (i.e., mass and annihilation channel) are identical for all dSphs, the sensitivity of detecting weak DM signal can be improved by a combined analysis. Such an analysis is also motivated by the similar (though very weak) signal in some dSphs. Here, we make the combined analysis for all the sample in Table \ref{tab:12dsph} and adopt the same approach as that in Ref. \cite{fermi14dsph,fermi15dsph,li16dsph,fermi2016dsph}. We create the combined likelihood function with the form of
\begin{equation}
\tilde{L}(\bm{\alpha},\{\bm{\theta}\})=\prod_{i}L_i(\bm{\alpha},\bm{\theta}_i)L_{{\rm J},i},
\label{eq:combinedlike}
\end{equation}
where $L_i$ is the likelihood given by Eq. (\ref{eq1}) for the $i$-th source and $L_{\rm J}$ is an additional term that is included in to take into account the uncertainty on the J-factor of each dSph, which is expressed as
\begin{equation}
L_{{\rm J},i}(J_i|J_{{\rm obs},i},{\sigma}_{i})=\frac{1}{\ln(10)J_{{\rm obs},i}\sqrt{2\pi}{\sigma}_{i}}
         \exp^{-[\log_{10}(J_{i})-\log_{10}(J_{{\rm obs},i})]^{2}/{2\sigma_{i}^2}}.
\end{equation}
The $J_{{\rm obs},i}$ and $\sigma_i$ are the measured J-factor and corresponding uncertainty for each dSph, while $J_i$ is the true value of the J-factor which is taken as nuisance parameters in the likelihood analysis.

The dSphs are dominated by dark matter, so their identification in optical is difficult. In addition, it is also challenging to identify a sufficient number of member stars in the dSphs and then determine the distribution of DM by stellar kinematics. The reliable J-factors for many dSph candidates are thus still unavailable. However, it is reported there exists an empirical relation between the heliocentric distances $d$ and J-factors of dwarf galaxies, which reads $J(d)\approx 10^{18.1\pm 0.1}(d/100~{\rm kpc})^{-2}$ \cite{Drlica-Wagner:2015xua}. 
In the combined analysis, we use the kinematically determined J-factors if available. Otherwise, the values estimated by the empirical relation are adopted. For the uncertainties on the estimated J-factors, we take a logarithmic symmetric value of $\pm0.4$ dex according to \cite{Drlica-Wagner:2015xua}. 
The estimated J-factors of the 12 dSphs and the spectroscopic J-factors from \cite{Geringer-Sameth:2014yza,Simon:2015fdw} are presented in Table \ref{tab:12dsph}.

\begin{figure}[!t]
\includegraphics[width=0.9\columnwidth]{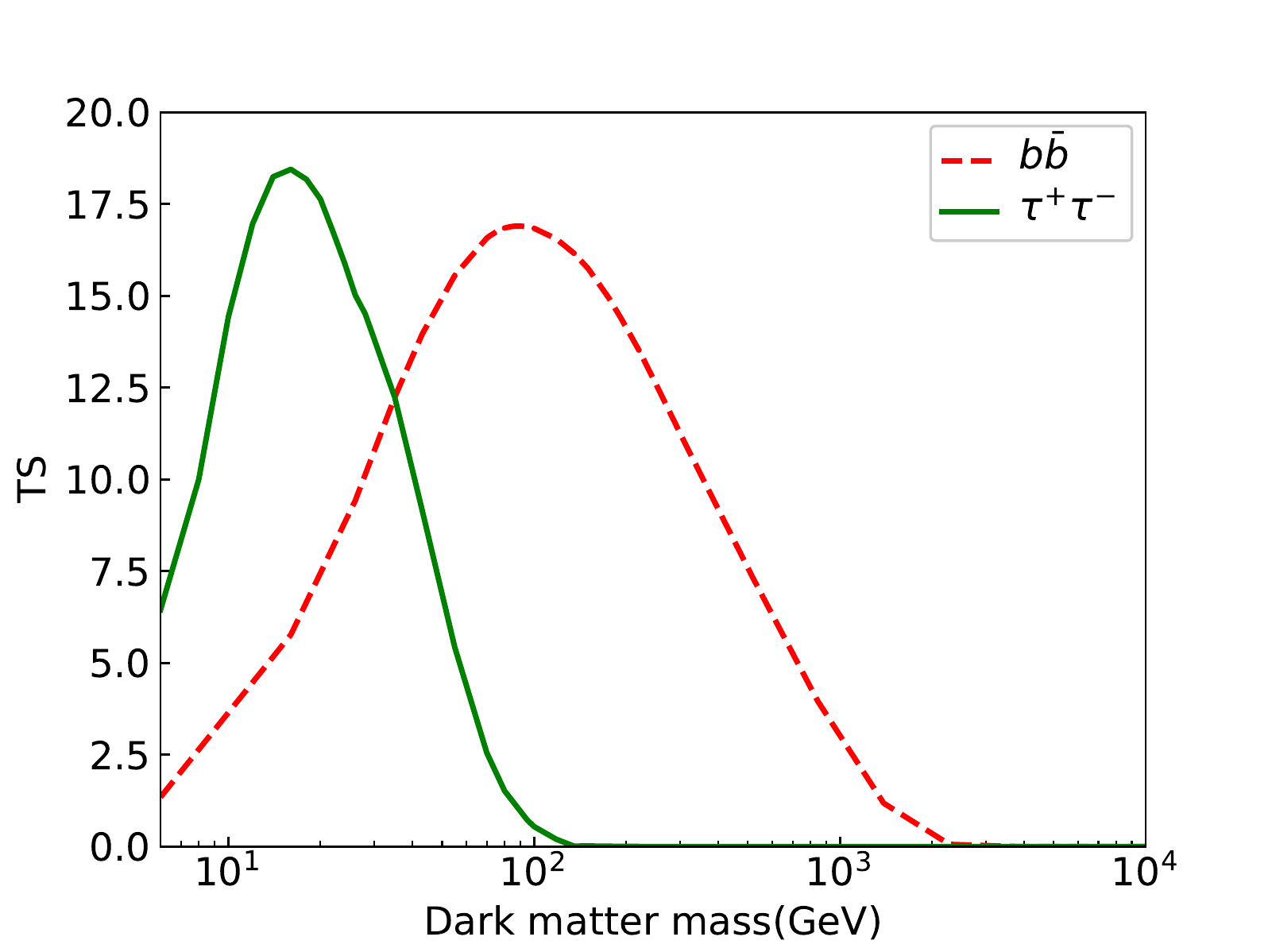}
\caption{The TS value of the tentative dark matter annihilation signal in the combined $\gamma$-ray analysis of the 12 dSph galaxies (or candidates) with distance $<50\,{\rm kpc}$.}
\label{fig:combinned}
\end{figure}

In Figure \ref{fig:combinned}, we show the TS values derived from a combined analysis as a function of the DM particle masses.
The largest excess found in the analysis is ${\rm TS} \approx 18.4$ at $m_{\chi}\approx 16\;{\rm GeV}$ for $\chi\chi\rightarrow \tau^{+}\tau^{-}$. In the case of $\chi\chi\rightarrow b\bar{b}$, the peak value of ${\rm TS} \approx 16.9$ is at $m_\chi \sim 90$ GeV.
The best-fit $m_\chi$ is very consistent with that in the analysis of Reticulum II, but the TS is larger.
We notice that the TS values obtained here are also the highest ones have ever been reported in the literature for the combined dSph analysis using Pass 8 data.
This may attribute to the longer data set used in our analysis which reflects that the combined TS is increasing as well.
The ${\rm TS} \approx 18.4\ (16.9)$ corresponds to a local significance of $4.3\sigma\ (4.1\sigma)$ considering one additional degree of freedom for the DM model.



\section{Summary and Discussion}
Among various objects, the Milky Way dwarf spheroidal galaxies are promising to identify the DM gamma-ray signal, due to the large DM content therein and the lack of high energy astrophysical processes. In this work, we revisist the analysis of 12 nearest dSphs (both confirmed and candidate) with 9 years of Fermi-LAT data. We don't find any statistically significant ($>5\sigma$) $\gamma$-ray excess in our sample.
However, in the direction of Reticulum II, a dSph that is $\sim 32$ kpc away, there is a weak excess with ${\rm TS}\approx 13.5$ for the annihilation channel $\chi\chi\rightarrow \tau^{+}\tau^{-}$  and the DM mass $m_\chi \approx 16$ GeV. For $\chi\chi\rightarrow b\bar{b}$, the largest TS value of $\sim10.9$ peaks at $m_\chi \sim 90$ GeV.

To improve the sensitivity of searching for DM signal, we have also performed a combined analysis of the 12 dSphs, which leads to a largest TS value of about 18.4 (16.9) at $m_{\chi}$ $\approx$ 16 GeV (90 GeV) for $\chi\chi\rightarrow \tau^{+}\tau^{-}$ ($b\bar{b}$) (see Fig.~\ref{fig:combinned}).

The weak gamma-ray emission from some dSphs have been reported before \cite{gs15ret2,hooper15ret2,Drlica-Wagner:2015xua,li16dsph,fermi2016dsph}. We notice that the peak TS values obtained in both the individual (Reticulum II) and combined analysis in this work have increased compared to the previous studies \cite{Drlica-Wagner:2015xua,li16dsph,fermi2016dsph}. In view of that we use a similar analysis method, the increase is likely owing to the longer (9 years) data set used.
By repeating the searching analysis for the 3, 6 and 9 years of Fermi-LAT data, we find that the peak $\rm TS$ of the $\gamma$-ray emission in the direction of Reticulum II is indeed growing with time (see Fig.~\ref{fig:3interval}). This results are in favor of the hypothesis that the gamma-ray excess from dSphs is a real signal (dark matter annihilation or a steady astrophysical source) rather than statistical fluctuation.

However, the tentative emission from dSphs is still weak at this time. The ${\rm TS} \approx 18.4\ (16.9)$ in the combined analysis corresponds to a local significance $\sim 4.3\sigma\ (4.1\sigma)$. The significance would be significantly lowered if we take into account the so-called trial-factor. Considering that we have selected 12 dSphs from totally $\sim50$ dSphs to do our analysis, the trial factor would be extremely large. But note that the 12 sources selected in this work are the nearest ones, which are most promising to emit detectable gamma-ray emission, thus it is reasonable to adopt a value of 50 for the trial factor related to the source selection. In conjunction with the trial factors sourced by the fact that we have scanned multiple DM masses and channels, the significance of the tentative emission from 12 dSphs reduces to $2.3\sigma$ and $2.6\sigma$ for $b\bar{b}$ and $\tau^{+}\tau^{-}$, respectively. Thus the excess is not statistically significant at present.
In addition, accounting for the non-poisson background of the analysis (due to the unmodeled point sources and inaccurate modeling of the diffuse background), the detection significance should be calibrated with blank-sky regions, which always leads to a much lower global significance \cite{fermi2016dsph}.


Nevertheless, detecting a tentative gamma-ray excess from some dSphs and especially finding it growing with time are encouraging. If coming from DM annihilation, the signal in the dSphs corresponds to a DM mass of $\sim 16\;{\rm GeV}$ for $\tau^{+}\tau^{-}$ (or $\sim 90\;{\rm GeV}$ for $b\bar{b}$), which is slightly higher than that needed to interpret the $\gamma$-ray excess associated with the Galactic center ($\sim25-70\,{\rm GeV}$ for $b\bar{b}$, $\sim8-15\,{\rm GeV}$ for $\tau^{+}\tau^{-}$, at $2\sigma$ confidence) \cite{Gordon:2013vta, Hooper:2013rwa, Daylan:2014rsa, Zhou:2014lva, Calore:2014xka} and that responsible for the low energy excess of antiprotons ($\sim20-80\,{\rm GeV}$ for $b\bar{b}$, at $2\sigma$ confidence) \cite{cui17antiproton,cuoco17antiproton}.
Even so, at this stage we can neither rule out the connection between them due to the low statistics of the tentative dSph emission.
The ongoing observations of Fermi-LAT, and some operating/future gamma-ray telescopes \cite{DAMPE:2017,gamma400,zhang14herd} will help to unveil the nature of this interesting (though weak) excess from dSphs.
\begin{acknowledgments}
We thank the anonymous referee for helpful suggestions and comments. This work is supported in part by the National Key Research and Development
Program of China (No. 2016YFA0400200), the National Natural Science Foundation
of China (Nos. 11525313, 11722328, 11773075, U1738210, U1738136).
\end{acknowledgments}
\bibliographystyle{apsrev4-1-lyf}
\bibliography{refs}
                                                                                     
\end{document}